\documentclass[runningheads]{llncs}
\RequirePackage{tikz}
\usepackage{graphicx}
\usepackage{amsmath}
\usepackage{amssymb}
\usepackage{bbm}
\usepackage{float}
\floatstyle{plaintop}
\restylefloat{table}
\begin{document}
%
\title{Fingerprints of data compression in EEG sequences \thanks{This work is part of FAPESP project 'Research, Innovation and Dissemination Center for Neuromathematics' (FAPESP grant 2013/07699-0), University of São Paulo project 'Mathematics, computation, language and the brain', project 'Plasticity in the brain after a brachial plexus lesion' (FAPERJ grant E26/010002474/2016) and project 'PROINFRA HOSPITALAR' (FINEP grant 18.569-8). Authors A.G and C.V. are partially supported by CNPq fellowships (grants 311 719/2016-3 and 309560/2017-9, respectively). Author C.D.V. is also partially supported by a FAPERJ fellowship (CNE grant E-26/202.785/2018). F.A.N. (grant number 88882.377124/2019-01) is supported by a PhD fellowship from the Brazilian Coordenação de Aperfeiçoamento de Pessoal de Nível Superior - CAPES}.}
%
%
\author{Fernando Araujo Najman\inst{1}\orcidID{0000-0001-9135-4397} \and
Antonio Galves\inst{1}\orcidID{0000-0001-8757-715X} \and
Claudia D. Vargas\inst{2}\orcidID{0000-0001-6202-3580}}
\authorrunning{F.A. Najman et al.}
%
\institute{Instituto de Matem\'atica e Estat\'istica, Universidade de S\~ao Paulo, Brazil 
\email{fernando.najman@usp.br}\\
\and
Instituto de Biof\'isica Carlos Chagas Filho, Universidade Federal do Rio de Janeiro, Brazil\\
}
\maketitle              
\begin{abstract}

 It has been classically conjectured that the brain compresses data by assigning probabilistic models to sequences of stimuli. An important issue associated to this conjecture is what class of models is used by the brain to perform its compression task. We address this issue by introducing a new statistical model selection procedure aiming to study the manner by which the brain performs data compression. Our procedure uses context tree models to represent sequences of stimuli and a new projective method for clustering EEG segments. The starting point is an experimental protocol in which EEG data is recorded while a participant is exposed to auditory stimuli generated by a stochastic chain. A simulation study using sequences of stimuli generated by two different context tree models with EEG segments generated by two distinct algorithms concludes this article.

\end{abstract}
\section{Introduction}\label{sec:int}
The conjecture that the brain compresses data by assigning probabilistic models to sequences of stimuli can be traced as early as the nineteenth century \cite{VonHelmholtz:67} (see for instance \cite{maheu_brain_2019}, \cite{rubin2016representation}, \cite{doya2007bayesian}, \cite{huang2011predictive}). Recently \cite{duarte_retrieving_2019} addressed this conjecture using the following combined probabilistic and experimental framework.
 EEG data is recorded while a participant is exposed to a sequence of auditory stimuli generated by a probabilistic algorithm. The question is whether his/her brain is able to identify statistical regularities in the sequence of stimuli and use them to compress the conveyed information. In this study, context tree models are used as a framework to represent the brain compression mechanism. This is a natural choice as any stationary stochastic chain of symbols with finite memory can be seen as a context tree model \cite{rissanen_universal_1983}. The rationale behind this approach is the following. Assume that the brain is able to identify the context at each step of the stimuli sequence and that this identification is expressed trough an EEG activity which differs from one context to another. The specificity of the EEG corresponding to each context means that they are independent realizations of the same probability measure on a suitable space of functions. In other terms, each context defines a different probability measure on the set of trajectories which can be realized by an EEG. If this is the case, encoding the sequence of EEG segments by symbols corresponding to the different probability measures produces a compressed stochastic chain conveying essentially the same information.
 
 To address this issue, \cite{duarte_retrieving_2019} proposes collecting together all EEG signals recorded after the last stimulus of any sequence of three consecutive acoustic stimuli. Then, using the projective method \cite{Cuesta2006}, together with a suitable variant of the Context Algorithm \cite{rissanen_universal_1983}, the equality in law of the EEG segments recorded after sequences ending with a common suffix is checked through a statistical test. If the test supports the null assumption of equality, the last element of the sequences with the same suffix is pruned. This procedure is repeated until the test rejects the null assumption. In conclusion, the tree of sequences obtained by the pruning procedure can be compared with the context tree generating the sequence of stimuli.
 
 Applying this methodology to experimental data, \cite{hernandez2020retrieving} found that the retrieved trees coincided with those generating the sequences in the pre-frontal cortex. A major drawback of this approach is the fact that the pruning procedure always produces a tree, since only the laws of the EEG segments recorded after sequences with a common suffix are compared. Therefore, the question about the class of models used by the brain to compress data remains open.
 
 To make a step forward, let us have a closer look at the structure of the sequences of stimuli considered in \cite{duarte_retrieving_2019}. These sequences assume values in $A = \{0, 1, 2\}$, where each symbol represents a distinct auditory stimulus. One of the sequences considered is the following. We start with the deterministic sequence
$$
\ldots 2 \ 1 \ 1 \ 2 \ 1 \ 1 \ 2 \ 1 \ 1 \ 2 \ldots \ .
$$
Then for each symbol $1$, we decide to either keep it with probability $(1-\epsilon)$ or to replace it by a $0$ with probability $\epsilon$, where $\epsilon \in [0, 1/2]$ is a fixed parameter. This choice is made independently at each symbol $1$. Let $(X_0, X_1, \ldots)$ be the resulting stochastic chain. 

This stochastic chain can be generated step by step by an algorithm using only information from the past. To generate $X_n$, we first look to the last symbol $X_{n-1}$.

\begin{itemize}
\item If $X_{n-1}=2$, then  
$$
X_n=\left\lbrace\begin{array}{ll}
1,& \mbox{with probability} \ 1- \epsilon,\\
0, & \mbox{with probability} \ \epsilon.\
\end{array}\right.
$$
\item If $X_{n-1}=1$ or $X_{n-1}=0$, then we need to go back one more step, 
\begin{itemize}
\item[$\centerdot$] if $X_{n-2}=2$, then
$$
X_n=\left\lbrace\begin{array}{ll}
1,& \mbox{with probability} \ 1- \epsilon,\\
0, & \mbox{with probability} \ \epsilon;\
\end{array}\right.
$$ 
\item[$\centerdot$]  if $X_{n-2}=1$ or $X_{n-2}=0$, then $X_n=2$ with probability $1$.
\end{itemize}
\end{itemize}

The algorithm described above is characterized by two elements:  
\begin{itemize}
    \item a partition $\tau$ of the set of all possible sequences of past units;  
    \item a family $p$ of transition probabilities indexed by the elements of $\tau$.
\end{itemize}
 The partition $\tau$ described above is given by 
 $$
 \tau = \{00, 10, 20, 01, 11, 21, 2\} \ .
 $$
 
 The pair $(\tau, p)$ having as first element a context tree and a second element a family of transition probabilities indexed by the contexts in $\tau$ is called a \textit{probabilistic context tree}. A stochastic chain $(X_n)_{n \geq 0}$ taking values in a finite alphabet $A$ and generated by a probabilistic context tree $(\tau, p)$, as described above, is called is a \textit{context tree model}.
 
 The second sequence of stimuli considered in \cite{duarte_retrieving_2019} is also a context tree model having 
 $$
 \tau = \{000, 100, 200, 10, 20, 01, 21, 2\} \ .
 $$
 The pair $(\tau, p)$ corresponding to this context tree model is presented in Figure \ref{fig:qua}. 
 
 Following \cite{rissanen_universal_1983}, we call \textit{context} any element of the partition $\tau$. Observe that the partition $\tau$ can be represented by a rooted and labeled \textit{tree}. Figure \ref{fig:ter} represents the context tree and the corresponding family of transition probabilities described above.

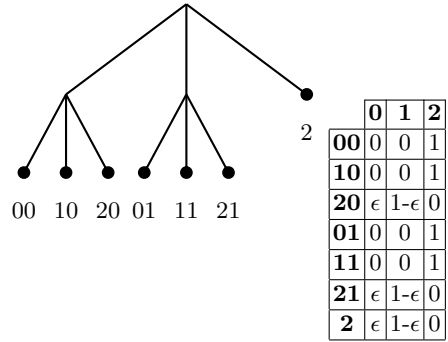
\begin{figure}[ht]
\centering
\begin{tikzpicture}[thick,scale=0.8]
     \tikzstyle{invisible} = [rectangle, node distance=0.5cm]
     \tikzstyle{level 1}=[level distance=1.5cm, sibling distance=2.0cm,]
     \tikzstyle{level 2}=[level distance=1.3cm, sibling distance=0.7cm]
    \coordinate
        child{{}
        	child {[fill] circle (2.5pt) node (00) {} }
        	child {[fill] circle (2.5pt) node (10) {} }
        	child {[fill] circle (2.5pt) node (20) {} }		        	
        	}
        child{ {} 
		child {[fill] circle (2.5pt) node (01) {} }
        	child {[fill] circle (2.5pt) node (11) {} }
        	child {[fill] circle (2.5pt) node(21) {} }        
        	}
        child{[fill] circle (2.5pt) node(2){} }
        ;
     \node [ invisible, below of=00 ]{00};
    \node [ invisible, below of=10 ]{10};
    \node [ invisible, below of=20 ]{20};
    \node [ invisible, below of=01 ]{01};
    \node [ invisible, below of=11 ]{11};
    \node [ invisible, below of=21 ]{21};
    \node [ invisible, below of=2 ]{2};
\end{tikzpicture}
\centering
\begin{tabular}{c|c|c|c|}
\cline{2-4}
                                   & \textbf{0} & \textbf{1} & \textbf{2} \\ \hline
\multicolumn{1}{|c|}{\textbf{00}} & 0          & 0          & 1          \\ \hline
\multicolumn{1}{|c|}{\textbf{10}} & 0          & 0          & 1          \\ \hline
\multicolumn{1}{|c|}{\textbf{20}} & $\epsilon$      & 1-$\epsilon$    & 0          \\ \hline
\multicolumn{1}{|c|}{\textbf{01}}  & 0      & 0    & 1          \\ \hline
\multicolumn{1}{|c|}{\textbf{11}}  & 0          & 0          & 1          \\ \hline
\multicolumn{1}{|c|}{\textbf{21}}  & $\epsilon$      & 1-$\epsilon$    & 0          \\ \hline
\multicolumn{1}{|c|}{\textbf{2}}  & $\epsilon$      & 1-$\epsilon$    & 0         \\ \hline
\end{tabular}

\caption{Context tree and family of probability measures corresponding to the model with context tree $\tau = \{00, 10, 20, 01, 11, 21, 2\}$ described in the example of section \ref{sec:int}.} 

\end{figure}\label{fig:ter}

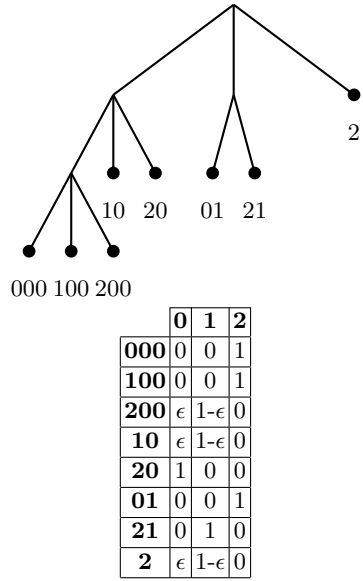
\begin{figure}[H]
\centering
\begin{tikzpicture}[thick,scale=0.8]
     \tikzstyle{level 1}=[level distance=1.5cm, sibling distance=2cm]
     \tikzstyle{level 2}=[level distance=1.3cm, sibling distance=0.7cm]
      \tikzstyle{level 3}=[level distance=1.3cm, sibling distance=0.7cm]
      \tikzstyle{invisible} = [rectangle, node distance=0.5cm]
    \coordinate
        child{{}
        	child { {}
        		child {[fill] circle (2.5pt) node(000){}} 
        		child {[fill] circle (2.5pt) node(100){}} 
        		child {[fill] circle (2.5pt) node(200){}} 
        		}
       		child {[fill] circle (2.5pt) node(10) {} }
        	child {[fill] circle (2.5pt) node(20) {} }		        	
        	}
        child{{} 
		child {[fill] circle (2.5pt) node(01){} }
        	child {[fill] circle (2.5pt) node(21){} }        
        }
        child{[fill] circle (2.5pt)  node(2){} }
    ;
    \node [ invisible, below of=000 ]{000};
    \node [ invisible, below of=100 ]{100};
    \node [ invisible, below of=200 ]{200};
    \node [ invisible, below of=10 ]{10};
    \node [ invisible, below of=20 ]{20};
    \node [ invisible, below of=01 ]{01};
    \node [ invisible, below of=21 ]{21};
    \node [ invisible, below of=2 ]{2};
\end{tikzpicture}

\centering
\begin{tabular}{c|c|c|c|}
\cline{2-4}
                                   & \textbf{0} & \textbf{1} & \textbf{2} \\ \hline
\multicolumn{1}{|c|}{\textbf{000}} & 0          & 0          & 1          \\ \hline
\multicolumn{1}{|c|}{\textbf{100}} & 0          & 0          & 1          \\ \hline
\multicolumn{1}{|c|}{\textbf{200}} & $\epsilon$      & 1-$\epsilon$    & 0          \\ \hline
\multicolumn{1}{|c|}{\textbf{10}} & $\epsilon$      & 1-$\epsilon$    & 0          \\ \hline
\multicolumn{1}{|c|}{\textbf{20}}  & 1          & 0          & 0          \\ \hline
\multicolumn{1}{|c|}{\textbf{01}}  & 0      & 0    & 1          \\ \hline
\multicolumn{1}{|c|}{\textbf{21}}  &  0     & 1    & 0         \\ \hline
\multicolumn{1}{|c|}{\textbf{2}}  & $\epsilon$      & 1-$\epsilon$    & 0         \\ \hline
\end{tabular}

\caption{Context tree and family of probability measures corresponding to the model with context tree $\tau = \{000, 100, 200, 10, 20, 01,  21, 2\}$ which is used in \cite{duarte_retrieving_2019}} 

\end{figure}\label{fig:qua}
  Given a context tree model, for each, $n \geq 0$, call $C_n$ the only context in $\tau$ which is a suffix of the sequence $(\ldots,X_{n-2},X_{n-1}, X_n)$. The examples considered in \cite{duarte_retrieving_2019} both have the following property: for each $n \geq 0$, the context $C_n$ is a suffix of the sequence obtained by the concatenation of the previous context $C_{n-1}$ and the next symbol $X_n$. This implies that the sequence $(C_n)_{n \geq 1}$ is a Markov chain of order 1 taking values in the context tree $\tau$. This property suggests an alternative procedure to treat the EEG data recorded using the experimental protocol employed in \cite{duarte_retrieving_2019}.
 
\section{Encoding a sequence of EEG segments as a Markov chain in $\tau$}\label{sec:enconding}

We start by introducing a general framework. Let $(X_n)_{n \geq 0}$ be a stochastic chain of stimuli generated by a probabilistic context tree $(\tau, p)$ as in the example described above. Let also $Y_n \in L^2([0, T])$ be the EEG segment recorded at a fixed electrode while a participant is exposed to stimuli $X_n$, where $T$ is the distance in time between two successive auditory stimuli onsets.  

Let us make the following assumptions. 
\begin{enumerate}
    \item The context tree $(\tau, p)$ is such that, for each $n \geq 0$, $C_n$ is a suffix of the sequence $C_{n-1}$ concatenated with $X_n$.
    \item The law of the EEG segment $Y_n$ is a function of the context $C_n$. We denote this law as $Q^{C_n}$.
    \item If $w$ and $w'$ are contexts belonging to $\tau$, and $Q^w$ and $Q^{w'}$ are the probability measures on $L^2([0, T])$ associated to $w$ and $w'$ respectively, then $Q^w = Q^{w'}$ if and only if $w = w'$.  
\end{enumerate}
 Let $\mathcal{Q}^{\tau}$ be the set of probability measures $\{Q^w : w \in \tau\}$. By the third assumption above, there is a one to one correspondence between $\mathcal{Q}^{\tau}$ and $\tau$. Ordering the contexts belonging to $\tau$ using the lexicographic order, we can represent the set of contexts $\tau$ by the set of positive integers $\{1, \ldots, |\tau|\}$. Let $I^*_n$ be the unique element in the set $\{1, \ldots, |\tau|\}$ corresponding to the law $Q^{C_n}$.

\begin{theorem}\label{Th:1}
Under the assumptions presented above, the sequence $(I^*_n)_{n \geq 0}$ is a Markov chain of order 1, taking values in the set $\{1, \ldots, |\tau|\}$.
\end{theorem}

If the sequence of stimuli and the corresponding sequence of EEG segments satisfy the assumptions presented above, then Theorem \ref{Th:1} suggests a new way to look at the sequence $(Y_n)_{n \geq 0}$ produced under the experimental protocol of \cite{duarte_retrieving_2019}.
 This is the content of the next section.
\section{Clustering the EEG segments}\label{sec:clusteringmeth}
 Take a positive integer $l \geq h(\tau)$, where $h(\tau)$ is the maximal length of the contexts belonging to $\tau$. Given a sample $((X_0, Y_0), \ldots, (X_N, Y_N))$ generated as described above, for all $u \in A^l$ let $\mathcal{Y}^u = \{Y_n : n = l-1, \ldots, n, \, \,  (X_{n-l+1},\ldots , X_{n}) = u\}$.
 
 If the EEG segments collected in $\mathcal{Y}^u$ have all been generated by the same probability measure $Q^u$ on $L^2([0, T])$, then the \textit{sample} $\mathcal{Y}^u$ can be used to approximate $Q^u$, and the approximation improves as the the length $N$ of the sample diverges. Therefore, the closeness in a suitable distance between two sets $\mathcal{Y}^u$ and $\mathcal{Y}^{u'}$, for different strings $u$ and $u'$, should give an indication of the closeness between the probability measures $Q^u$ and $Q^{u'}$. We now implement this idea using the projective method introduced in \cite{Cuesta2006}.
 
We start by generating a realization $B = (B(t) : t \in [0, T])$ of the Brownian Bridge. Then we project all the EEG segments $(Y_0, \ldots, Y_N)$ using this fixed realization of the Brownian Bridge. This is done as follows. For every $n = 0, \ldots, N$, the projection of $Y_n$ in the direction $B$ is obtained by the internal product in the Hilbert space $L^2([0, T])$
$$
R_{n, B} = \int_{0}^{T}B(t)Y_n(t)dt \ .
$$

For each $u \in A^l$, the projection in the direction $B$ of the set $\mathcal{Y}^u$ is naturally defined as
$$
\mathcal{Y}_B^u = \{R_{n, B}: Y_n \in \mathcal{Y}^u\}
$$
Let $F_B^u$ be the empirical distribution obtained from the sample of real numbers $\mathcal{Y}^u_B$
$$
F_B^u(t) = \frac{1}{|\mathcal{Y}_B^u|}\sum_{y_B^u \in \mathcal{Y}_B^u}\mathbbm{1}_{\{y_B^u \leq t\}} \, \, , t \in \mathbb{R}
$$

Then, for each pair of sequences $u$ and $v$ in $A^l$, we define the renormalized Kolmogorov-Smirnov distance between the empirical measures obtained from the samples $\mathcal{Y}^u_B$ and $\mathcal{Y}^v_B$
 $$KS(\mathcal{Y}_B^u, \mathcal{Y}_B^v) = \sqrt{\frac{|\mathcal{Y}_B^u| |\mathcal{Y}_B^v|}{|\mathcal{Y}_B^u|+ |\mathcal{Y}_B^v|}}\sup_{t \in \mathbb{R}}\{|\int^t_{-\infty}[F_B^u(x)-F_B^v(x)]dx|\}.$$ 

Let $B_1, \ldots, B_M$ be independent copies of the Brownian Bridge. We define the distance between $\mathcal{Y}^u$ and $\mathcal{Y}^v$ as
$$
D_M(\mathcal{Y}^u, \mathcal{Y}^v) = \frac{1}{M}\sum_{i = 1}^M KS(\mathcal{Y}_{B_i}^u, \mathcal{Y}_{B_i}^v)
$$

For details on the projective method we refer the reader to  \cite{duarte_retrieving_2019}.
 
 For $k = 2, \ldots, |A|^l$ we want to partition the set $\{\mathcal{Y}^u : u \in A^l\}$ as follows. We start by choosing arbitrarily $k$ different sequences $u^0_1, \ldots, u^0_k$ such that, for all $i$ and $j$ with $i \neq j$, $u^0_i \neq u^0_j$. These sequences will be used as medoids of the first candidate partition defined as follows. $\mathcal{P}^{u^0_j}$ is the set of all $v \in A^l$ such that 
 $$
 D_M(\mathcal{Y}^{u^0_j}, \mathcal{Y}^{v}) = \min \{D_M(\mathcal{Y}^{u^0_i}, \mathcal{Y}^{v}) : i = 1, \ldots, k \} \ .
 $$
 Now we iterate the procedure. For each all $v \in \mathcal{P}^{u^0_j}$ and $j = 1, \ldots, k$, we define 
$$
u^1_j = \arg\min \{\sum_{v \neq v'}D_M(\mathcal{Y}^{v}, \mathcal{Y}^{v'})\} \ ,
$$
and then define $\mathcal{P}^{u^1_j}$ is the set of all $v \in A^l$ such that 
 $$
 D_M(\mathcal{Y}^{u^1_j}, \mathcal{Y}^{v}) = \min \{D_M(\mathcal{Y}^{u^1_i}, \mathcal{Y}^{v}) : i = 1, \ldots, k \} \ .
 $$
 The partition $\mathcal{C}^k = \{\mathcal{C}^k_1, \ldots, \mathcal{C}^k_k\}$ is the limit of the above described procedure. For more details on the algorithm we refer the reader to \cite{park2009simple}.
 
 We can now use the partition $\mathcal{C}^k$ to encode the sequence of EEG segments $(Y_1, \ldots, Y_N)$ as follows. Let $I_n^k$ be the index of the cluster containing the EEG segment $Y_n$. When $k = |\tau|$, Theorem \ref{Th:1} predicts that the sequence $(I_n^k)_{n \geq 0}$ is a Markov chain with memory of order 1. In the next section we address this prediction using a simulation study.

\section{Simulating the samples}\label{sec:simula}
To discuss the performance of the proposed method we conducted a simulation study following the assumptions in sections \ref{sec:enconding} and \ref{sec:clusteringmeth}, using different values for the length $N = 100, 600, 900$ of the sequence of the simulated sample $((X_1,Y_1), \ldots, (X_N,Y_N))$.

The sequence of stimuli were generated by the two context tree models presented in Section \ref{sec:int}, henceforth called $(\tau_1, p_1)$ and $(\tau_2, p_2)$ respectively, with $\epsilon = 0.2$.

We used two independent and distinct algorithms to simulate the sequence of EEG segments generated as function of the stimuli contexts. This was done to exemplify the independence of the methodology presented here with to the specific probability measures generating the EEG segments. Both EEG generating algorithms simulated the EEG segments as ordered vectors containing 250 real numbers each. 

\begin{figure}[h]
    \centering
    \includegraphics[width=1
    \textwidth]{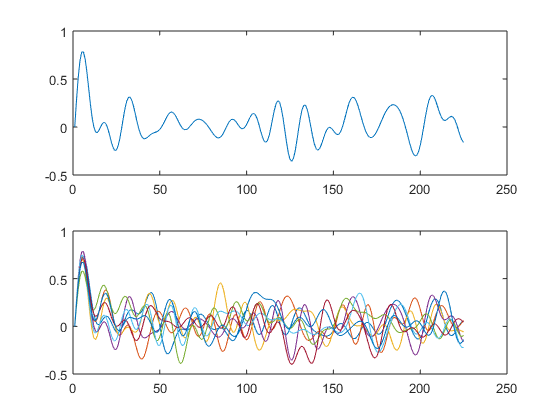}
    \caption{The upper image presents one of the random averages $S^w$ for a fixed $w \in \tau_2$ used in the Average Model Algorithm EEG simulation algorithm. The lower image presents all the random averages $S^w : w \in \tau^2$ used in the same simulation.}
    \label{fig:avgs}
\end{figure}

The first algorithm used to generate the EEG vector, henceforth called \textbf{Average Model Algorithm}, can be described as follows.
\begin{enumerate}
    \item  For each context $w \in \tau$, choose independently and uniformly 40 real numbers $s^w = \{s_1^w, \ldots, s_{40}^w\}$ in the interval $[0, 40]$
    \item  For each sorted number, define the discretized sine wave with frequency $s_i^w$ 
    $$
    \phi_i^w(t) = \sin(\frac{t}{250}2 \pi s_i^w ) \ .
    $$
    \item For each context, compute the average vector 
    $$
    S^w(t) = \frac{1}{40}\sum_{i = 1}^{40}\phi^w_i(t)
    $$
    \item For $n  = h(\tau)-1, \ldots, N$ simulate the EEG segment as follows
    $$
    Y_n(t) = S^{C_n}(t)+\xi(t)
    $$
    where $C_n$ is the context present at step $n$, and $\xi(s) : s = 1, \ldots 250$ are independent random variables with normal distribution of mean $0$ and variance $0.3$. 
\end{enumerate}
Figure \ref{fig:avgs} shows one particular simulation of the set $\{S^w : w \in \tau_2\}$.

The second algorithm used to generate the EEG vector, henceforth called \textbf{Auto-Regressive Algorithm}, can be described as follows.
\begin{itemize} 

\item For each context $w \in \tau$, choose independently and uniformly 2 real numbers $\theta^w = \{\theta^w_1, \theta^w_1\}$ in the interval $[-1, 1]$,
\item For all $n = 1, \ldots, N$, let $Y_n(1) = 0$.
\item For each $n = h(\tau), \ldots, N$ and for each $t = 1, \ldots, N$ simulate the EEG segment as follows
$$
Y_n(t) = \theta^{C_n}_1+\theta^{C_n}_2Y_{n}(t-1)+\xi_t \ ,
$$
where $C_n$ is the context present at step $n$, and $\xi(s) : s = 1, \ldots 250$ are independent random variables with normal distribution of mean $0$ and variance $0.05$. 
\end{itemize}

\begin{figure}[h]
    \centering
    \includegraphics[width=1
    \textwidth]{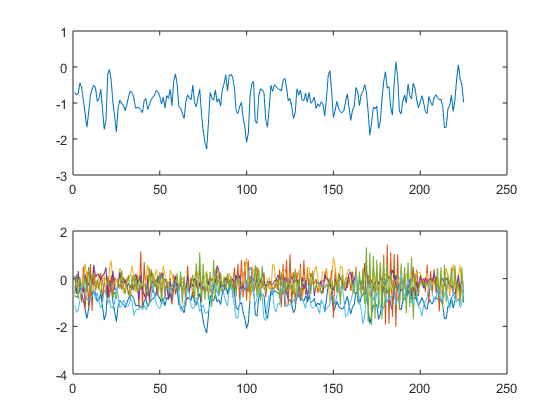}
    \caption{The upper image presents one of the random averages $S^w$ for a fixed $w \in \tau_2$ used in the Average Model Algorithm EEG simulation algorithm. The lower image presents all the random averages $S^w : w \in \tau^2$ used in the same simulation.}
    \label{fig:ars}
\end{figure}

Figure \ref{fig:ars} shows examples of the simulated EEG segments obtained using the Auto-Regressive Algorithm.

For each sequence length $N = 300, 600, 900$, each probabilistic context tree and each EEG segment generator we made 100 simulations of samples of length $N$. Each simulation was made using a new set of random parameters chosen independently.

The clustering of the data was done with the procedure described in Section \ref{sec:clusteringmeth}. For each sample we used $M = 5000$ independently generated Brownian Bridges to perform the projections required to define $D_M$.

Let $(I^{7}_n)_{n \geq h(\tau_1)}$ and $(I^{8}_n)_{n \geq h(\tau_2)}$ be the index sequences obtained by the method described above for $\tau = \tau_1$ or $\tau = \tau_2$, respectively. To identify the order of the chains $(I^{7}_n)_{n \geq h(\tau_1)}$ and $(I^{8}_n)_{n \geq h(\tau_2)}$ we used a slightly modified version of the statistical model selection procedure SMC introduced in \cite{galves2012context}. All the codes used for the simulations, including the version of the SMC procedure we implemented are available at '\url{https://github.com/FernandoNajman/Fingerprints-of-data-compression-in-EEG-sequences}'.

\section{Results and discussion}

     As predicted by Theorem \ref{Th:1}, we observed that for the majority of simulated samples generated by the two probabilistic context trees $(\tau_1, p_1)$ and $(\tau_2, p_2)$ and the two EEG generator algorithms, the SMC procedure identified the retrieved sequences of cluster indexes $(I^{7}_n : n = h(\tau_1), \ldots, N)$ and $(I^{8}_n : n = h(\tau_2), \ldots, N)$  as Markov chains of order 1. These results are summarized in Tables 1, 2, 3 and 4.
    
    The results obtained from the simulations employing the two probabilistic context trees and the two EEG generators are very similar.
    
     Finally, in the simulation study it appeared that even with the smallest value of $N = 300$, we have around $80 \%$ of the obtained sequences of indexes identified as order 1 Markov chains by the SMC procedure. 
     
     Furthermore, as expected, for each probabilistic context tree and each EEG generator algorithm, the number of simulations in which the SMC procedure identified the sequences of cluster indexes as Markov chains of order 1 increases, when the length $N$ of the sample increases. This was expected as a consequence of the consistency of the SMC procedure \cite{galves2012context}. This is also an indication of the accuracy of the method introduced in the present article to identify the different probability measures generating the EEG segments.
    
In conclusion this article presented a new statistical model selection procedure aiming to identify the manner by which the brain performs data compression. It also introduced a new method for clustering functional data which can find use beyond the original neurobiological motivation.

\begin{table}[H]
\centering
\begin{tabular}{c|c|c|c|c|c|}
\cline{2-5}
                               & $\mathcal{N}(1)$ & $\mathcal{N}(2)$ & $\mathcal{N}(3)$ & $\mathcal{N}(4)$ \\ \hline
\multicolumn{1}{|c|}{N = 300}         & 83       & 9       &   7      & 1        \\ \hline
\multicolumn{1}{|c|}{N = 600}         & 91       & 8        & 1        & 0        \\ \hline
\multicolumn{1}{|c|}{N = 900}         & 100       & 0        & 0        & 0        \\ \hline
\end{tabular}
\caption{Simulation results. For each $N = 300, 600, 900$ one hundred samples of length $N$ were simulated with sequences of stimuli generated by $(\tau_1, p_1)$ and the corresponding EEG segments were produced using the Average Model Algorithm. $\mathcal{N}(1)$, $\mathcal{N}(2)$, $\mathcal{N}(3)$ and $\mathcal{N}(4)$ indicate the number of times the statistical selection procedure SMC assigned memory 1, 2, 3 and 4, respectively, to the simulated samples.}
\end{table}\label{tab:resul_sim_avgs}

\begin{table}[H]
\centering
\begin{tabular}{c|c|c|c|c|c|}
\cline{2-5}
                               & $\mathcal{N}(1)$ & $\mathcal{N}(2)$ & $\mathcal{N}(3)$ & $\mathcal{N}(4)$ \\ \hline
\multicolumn{1}{|c|}{N = 300}         &     85   &  12      &  1       & 2         \\ \hline
\multicolumn{1}{|c|}{N = 600}         &     93   &  4       & 3        & 0         \\ \hline
\multicolumn{1}{|c|}{N = 900}         &     96   &  4       & 0        & 0        \\ \hline
\end{tabular}
\caption{Simulation results. For each $N = 300, 600, 900$, one hundred samples of length $N$ were simulated with sequences of stimuli generated by $(\tau_1, p_1)$ and the corresponding EEG segments were produced using the Auto-Regressive Algorithm. $\mathcal{N}(1)$, $\mathcal{N}(2)$, $\mathcal{N}(3)$ and $\mathcal{N}(4)$ indicate the number of times the statistical selection procedure SMC assigned memory 1, 2, 3 and 4, respectively, to the simulated samples.}
\end{table}\label{tab:resul_sim_ar}

\begin{table}[H]
\centering
\begin{tabular}{c|c|c|c|c|c|}
\cline{2-5}
                               & $\mathcal{N}(1)$ & $\mathcal{N}(2)$ & $\mathcal{N}(3)$ & $\mathcal{N}(4)$ \\ \hline
\multicolumn{1}{|c|}{N = 300}         & 77       & 13       & 8        & 2        \\ \hline
\multicolumn{1}{|c|}{N = 600}         & 87       & 11        & 0        & 2        \\ \hline
\multicolumn{1}{|c|}{N = 900}         & 93       & 7        & 0        & 0        \\ \hline
\end{tabular}
\caption{Simulation results. For each $N = 300, 600, 900$, one hundred samples of length $N$ were simulated with sequences of stimuli generated by $(\tau_2, p_2)$ and the corresponding EEG segments were produced using the Average Model Algorithm. $\mathcal{N}(1)$, $\mathcal{N}(2)$, $\mathcal{N}(3)$ and $\mathcal{N}(4)$ indicate the number of times the statistical selection procedure SMC assigned memory 1, 2, 3 and 4, respectively, to the simulated samples.}
\end{table}\label{tab:resul_sim_avgs2}

\begin{table}[H]
\centering
\begin{tabular}{c|c|c|c|c|c|}
\cline{2-5}
                               & $\mathcal{N}(1)$ & $\mathcal{N}(2)$ & $\mathcal{N}(3)$ & $\mathcal{N}(4)$ \\ \hline
\multicolumn{1}{|c|}{N = 300}         &     87   &  11      &  1       & 1         \\ \hline
\multicolumn{1}{|c|}{N = 600}         &     93   &  6       & 1        & 0         \\ \hline
\multicolumn{1}{|c|}{N = 900}         &     94   &  6       & 0        & 0        \\ \hline
\end{tabular}
\caption{Simulation results. For each $N = 300, 600, 900$, one hundred samples of length $N$ were simulated with sequences of stimuli generated by $(\tau_2, p_2)$ and the corresponding EEG segments were produced using the Auto-Regressive Algorithm. $\mathcal{N}(1)$, $\mathcal{N}(2)$, $\mathcal{N}(3)$ and $\mathcal{N}(4)$ indicate the number of times the statistical selection procedure SMC assigned memory 1, 2, 3 and 4, respectively, to the simulated samples.}
\end{table}\label{tab:resul_sim_ar2}

 \bibliographystyle{splncs04}
 \bibliography{Fingerprints_of_data_compression_in_EEG_sequences}

\end{document}